\documentclass[showkeys,twocolumn,superscriptaddress]{revtex4-1}
\usepackage{graphicx}
\usepackage{amsmath}

\begin{document}

\title{The dynamics of the angular and radial density correlation scaling exponents in fractal to non-fractal morphodynamics}

\author{J. R. Nicol\'as-Carlock}
\email{jnicolas@ifuap.buap.mx}
\thanks{author to whom correspondence should be addressed.}
\affiliation{Instituto de F\'isica, Benem\'erita Universidad Aut\'onoma de Puebla, Apdo. Postal. J-48, Puebla 72570, M\'exico.}

\author{J. M. Solano-Altamirano}
\affiliation{Facultad de Ciencias Qu\'imicas, Benem\'erita Universidad Aut\'onoma de Puebla, Apdo. Postal. J-48, Puebla 72570, M\'exico.}

\author{J. L. Carrillo-Estrada}
\affiliation{Instituto de F\'isica, Benem\'erita Universidad Aut\'onoma de Puebla, Apdo. Postal. J-48, Puebla 72570, M\'exico.}

\begin{abstract} 
Fractal/non-fractal morphological transitions allow for the systematic study of the physics behind fractal morphogenesis in nature. In these systems, the fractal dimension is considered a non-thermal order parameter, commonly and equivalently computed from the scaling of the two-point radial- or angular-density correlations. However, these two quantities lead to discrepancies during the analysis of basic systems, such as in the diffusion-limited aggregation fractal. Hence, the corresponding clarification regarding the limits of the radial/angular scaling equivalence is needed. In this work, considering three fundamental fractal/non-fractal transitions in two dimensions, we show that the unavoidable emergence of growth anisotropies is responsible for the breaking-down of the radial/angular equivalence. Specifically, we show that the angular scaling behaves as a critical power-law, whereas the radial scaling as an exponential that, under the fractal dimension interpretation, resemble first- and second-order transitions, respectively. Remarkably, these and previous results can be unified under a single fractal dimensionality equation. 
\end{abstract}

\keywords{morphology, scaling, fractality, transitions, universality}

\maketitle


\section{Introduction}

In his celebrated book, ``On Growth and Form'', D'arcy Thompson suggested that natural selection was not the only factor shaping the biological development of species, but that in nature, ``no organic forms exist save such as are in conformity with physical and mathematical laws'' \cite{thompson1942,pball2013}. Since then, some of the most important scientific endeavours of the modern era have dealt with the exploration of the fundamental physical processes behind morphogenesis, the establishment of the mathematical tools for their analysis, and the development of appropriate control mechanisms for their further scientific and technological application \cite{benjacob1997,tallinen2016}. As complex as this problem is, the study of diverse systems using the concepts and tools of fractal geometry, dynamical systems, and out-of-equilibrium physics, along with the development of \textsl{ad-hoc} computational models, has led to the discovery of fundamental pattern-formation mechanisms \cite{benjacob1990,vicsek1992book,meakin1998book}. Among these, Laplacian growth stands as one of the most remarkable, given it's capability to reproduce the intricate fractal patterns observed in seemingly unrelated living and non-living systems (such as bacterial colonies, mineral deposition, viscous fingering, and dielectric breakdown) \cite{benjacob1997,vicsek2001book,mathiesen2006,sander2011}. Even with relevant applications in current neuroscience and cancer research \cite{sturmberg2013,lennon2015,diieva2016}. However, a similar universality or unification in terms of a general analytical framework for the study of the fractal morphodynamics of these systems has been elusive \cite{meakin1998book,sander2011}.

\begin{figure}[t!]
\includegraphics[width=3.4in]{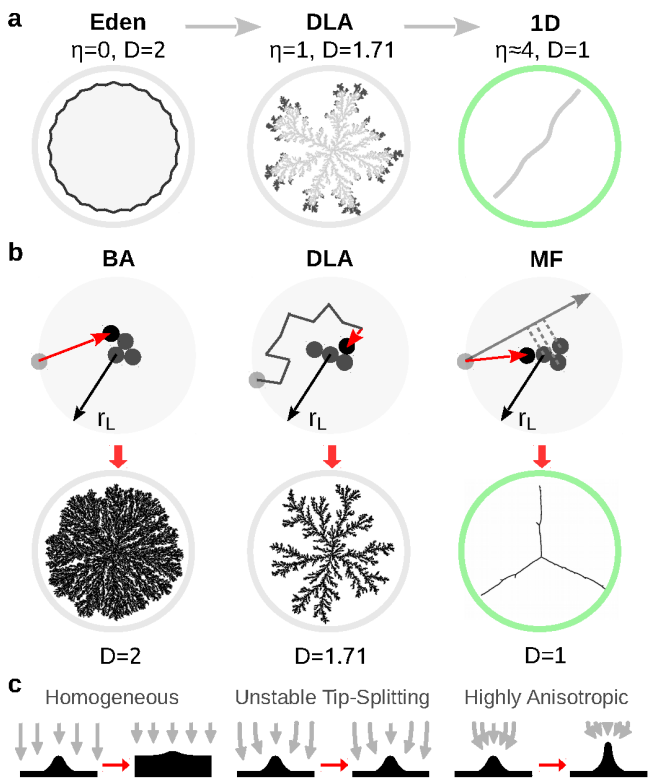}
\caption{\label{fig1}\textbf{Models.} (a) Two-dimensional Laplacian or DBM clusters for characteristic values of the parameter, $\eta$, and corresponding fractal dimension, $D$. (b) Clusters generated by the BA, DLA, and MF particle-aggregation models. In BA, particles follow straight-line trajectories before aggregating on contact to an initial seed particle; in DLA, particles follow random-walk trajectories, and in MF, particles are immediately aggregated to the closest one along its direction of motion due to an infinite attractive interaction. Here, $r_L(\gg 1)$ is the launching radius circle. (c) Sketch of the corresponding growth dynamics for each column. Arrows point towards the preferential growth directions. } 
\end{figure}

To understand this problem, let us first recall that in the Laplacian framework -- as originally introduced in the dielectric breakdown model (DBM) \cite{niemeyer1984,hayakawa1986,sanchez1993, nicolas2019} -- the growth is controlled \textit{via} a growth-probability distribution, $\sigma\propto|\nabla\phi|^\eta$. Here $\phi$ is a scalar field associated to the potential energy landscape of the growing surface, and $\eta$ is a positive real parameter associated to non-linear effects. As $\eta$ changes from zero to infinity, different structures emerge for a given set of initial boundary conditions. For example, considering an initial seed-point in two dimensions, different values of $\eta$ generate either homogeneous growth of compact clusters ($\eta=0$), unstable growth of dendritic-shaped fractals ($\eta\sim1$), or highly anisotropic growth of linear clusters ($\eta\approx4$) \cite{sander2011, nicolas2019} (see Fig.~1a). Hence, as a function of the control parameter $\eta$, these structures define what it is refer to as a continuous morphological transition. 

While these DBM clusters define one of the most fundamental morphological transitions, they are not unique. In fact, the identification of the most basic growth mechanisms that give origin to fractal clusters has lead to the generation of diverse fractal/non-fractal morphological transitions \cite{nicolas2016,nicolas2017}. For instance, the combination of the stochastic growth dynamics of the fundamental diffusion-limited aggregation (DLA) \cite{witten81} or ballistic aggregation (BA) \cite{vold63} models, with the highly energetic growth dynamics of the mean-field (MF) aggregation model (see Fig.~1b), have led to the creation of the non-trivial BA-MF and DLA-MF morphological transitions \cite{nicolas2017}. In these systems, the transition is controlled through the probability of aggregation under MF dynamics or mixing parameter, $p$, that continuously goes from $p=0$ (BA/DLA) to $p=1$ (MF), see Fig.~2. 

Generally, given the complex morphodynamics of these systems, most characterization approaches rely on the numerical estimation of the fractal dimension, $D$, as a function of system-specific growth parameters (e.g. $\eta$ or $p$). So far, there is no theoretical or analytical approach able to unify the numerical results obtained through diverse methodologies into a single framework. For example, one standard characterization method consists on the estimation of the fractal dimension from the scaling of quantities such as the two-point radial-density correlation, $C(r)$ \cite{meakin1998book}. This correlation is known to behave as $C(r)\sim r^{-\alpha_r}$, and the fractal dimension is obtained from $D_r=d-\alpha_r$, with $d$ being the Euclidean dimension of the embedding space. In a similar approach, the two-point angular-density correlation function, $C_R(\theta)$, is regarded as an equivalent method to estimate the fractal dimension of these structures \cite{meakin1985,alves2006,braga2017}. For $\theta\ \ll 1$, the angular correlation is also known to behave as $C_R(\theta)\sim \theta^{-\alpha_\theta}$, and the fractal dimension is obtained from $D_{\theta}=d-\alpha_\theta$, with $R$ being a distance relative to a certain origin \cite{meakin1985,alves2006}. Clearly, both expressions for the angular/radial scaling indicate that $\alpha_r=\alpha_\theta$ (or $D_r=D_\theta$). However, significant discrepancies \cite{meakin1985,alves2006,kolb1985,arneodo1992,mandelbrot1995,mandelbrot2002} have been observed in the fundamental two-dimensional DLA model (or DBM with $\eta=1$). In particular, the angular scaling is consistently reported as $\alpha_\theta=0.41$ \cite{meakin1985,alves2006}, leading to $D_{\theta}=1.59$, whereas the radial measurement is reported as, $\alpha_r\approx 0.29$, which leads to $D_r=1.71$, the accepted fractal dimension for this model \cite{sander2011}. 

\begin{figure}[t!]
\includegraphics[width=3.4in]{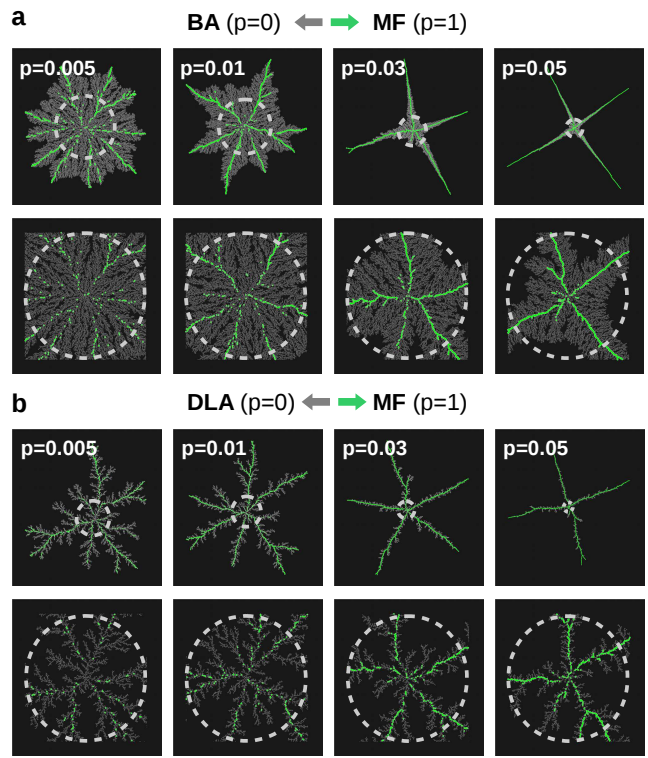}
\caption{\label{fig2}\textbf{BA/DLA-MF transitions.} Typical clusters formed in the (a) BA-MF and (b) DLA-MF morphological transitions, for the indicated values of the mixing parameter, $p$. Aggregated particles that followed the MF dynamics are coloured in green. The inside of the dashed circle (of radius $R=200$ particle diameters) is the region of analysis in this study. For further details, see Methods.} 
\end{figure}

Therefore, in this work we study the limits of this radial/angular equivalence in order to provide some insight into the origin of the previously observed discrepancy and the possibility of a unified and consistent radial/angular description. To this end, we make use of a recently introduced mathematical model for the fractal dimension, $D$, of structures that undergo a continuous geometrical (symmetry-breaking) transformation as function of an effective control parameter, $\Phi$ \cite{nicolas2017}. This model relies on a single dimensionality function, $D(D_0,\Phi)$, given by,
\begin{align}
\label{eq01}
D(D_0,\Phi)=1+(D_0-1)e^{-\Phi}.
\end{align}
Here, $D_0(\leq d)$ is the dimension of the initial configuration, and $\Phi$ is a function that encodes all the information regarding the order/disorder elements of the growth dynamics. It has the form,
\begin{align}
\label{eq02}
\Phi(D_0,\zeta)=\Phi_0 \zeta ^\chi,
\end{align}
where, $\zeta$ is the control parameter of the transition (e.g. $\eta$ or $p$), and $\Phi_0=\Lambda/D_0$, with $\Lambda$ and $\chi$ being two characteristic parameters associated to the amplitude and steepness of the scaling dynamics, respectively. 

Our analysis is focused on three examples: the BA-MF, DLA-MF, and the DBM transitions. Special attention is paid to the analytical treatment that $D(D_0,\Phi)$ provides to the angular fractal dimensions as a function of the corresponding control parameters. These results are compared to the radial scaling description (obtained in a previous analysis \cite{nicolas2017}) in order to determine their degree of equivalence and complementarity. In particular, we also look into their levels of predictability, this is, the extent to which the knowledge of just one of the radial or angular measurements predicts the other one. Afterwards, the role of the fractal dimension as an order parameter is treated. Finally, a complete scaling or fractality framework (which incorporates additional radius of gyration and mean-field results) is presented, revealing remarkable universal features.

\section{Methods}

\subsection{Growth models} 

As described in previous work \cite{nicolas2017}, in all simulations, each particle has a diameter equal to one (the basic unit of distance of the system). For BA or MF (see Fig 1b), we follow a standard procedure in which particles are launched at random from the circumference of a circle of radius $r_L = 2r_{max} + \delta$, with equal probability in position and direction of motion. Here, $r_{max}$ is the distance of the farthest particle in the cluster with respect to the seed particle placed at the origin. In our simulations we used $\delta=1000$ particle diameters to avoid undesired screening effects. For MF, particles always aggregate to the closest particle in the cluster with respect to their incoming path. This is determined by the projected position of the aggregated particles along the direction of motion of the incoming particle (see Fig 1b). In the case of DLA, particles were launched from a circumference of radius $r_L = r_{max} + \delta$, with $\delta=100$. The mean free path of the particles is set to one particle diameter. We also used a standard scheme that modifies the mean free path of the particles as they wander at a distance larger than $r_L$ or in-between branches, as well as the common practice of  setting a killing radius at $r_K = 2r_L$ in order to speed up the aggregation process. 

In order to mix different aggregation dynamics, a Monte Carlo scheme of aggregation is implemented using the BA, DLA and MF models. The combination between pairs of models results in the DLA-MF and BA-MF transitions, controlled by the mixing parameter $p\in[0,1]$, associated with the probability or fraction of particles aggregated under MF dynamics, $p=N_\mathrm{MF}/N$, where $N$ is total number of particles in the cluster. Therefore, as $p$ varies from $p=0$ (pure stochastic dynamics given by the BA or DLA models) to $p=1$ (purely energetic dynamics given by the MF model), it generates the two transitions under analysis (Fig. 2). The evaluation of the aggregation scheme to be used is only updated once a particle has been successfully aggregated to the cluster under such dynamics.

\subsection{Angular correlation and scaling analysis} 

To measure the angular correlation we followed the definition \cite{meakin1985}:
\begin{equation}
\nonumber
   C_R(\theta)=\frac{1}{N}\sum_{n=1}^{K/2}\rho_R(\theta)\rho_R(\theta+n\delta\theta).
\end{equation}
Here $\rho_R(\theta)$ is the number of particles contained in a box of size $\delta R\delta\theta$, centred at $(R,\theta)$, and $N$ is the size of the cluster. For the data shown in Figs.~3a and 3b, we used $RdRd\theta\approx \mathcal{N}\pi/4$. The size of the counting box is evaluated from the relation $\mathcal{N}\pi/4=fR^2\cdot2\pi/K$, hence $K=8fR^2/\mathcal{N}$.
Here, $\mathcal{N}$ is some number of particles (determined heuristically), $K$ determines $d\theta$ ($d\theta=2\pi/K$), and $f$ sets $dR$ ($dR=fR$). For the calculations presented in this work, we used $\mathcal{N}=3$ and $f=0.1$ for computing $c_R(\theta)$, $R=7,9,11,13,15$; $\mathcal{N}=10$ and $f=0.1$ for computing
$C_R(\theta)$, $R=20,40,60,80$ and $\mathcal{N}=25$; and $f=0.05$ for computing $C_R(\theta)$, $R=100,125,150,175,200$. In Figs.~3a and 3b, each curve is the average of 128 clusters,  \textit{i.e.}, we used 128 clusters for each value of p (16 different values of the latter), and for each growth dynamics (either BA/MF or DLA/MB), which accounts for 4096 clusters in total. Here, we want to remark that all our measurements were performed upon fully grown clusters, that is, clusters of $1.5\times10^{5}$ particles with a size of $R\geq366$ particle diameters and radial distances of $r\leq R+\delta R = 200+20$ in order to avoid undesirable border effects.

Regarding the angular scaling analysis (Figs.~3c and 3d), the angular scaling exponent, $\alpha_{\theta}(p,R)$, is measured for different values of the control parameter $p\in[0,1]$, and for increasing values of $R$. Here each dot is extracted from the slopes of linear fits according to $C_R(\theta)\sim\theta^{-\alpha_\theta}$, for $\theta\ll 1$ (as sketched in Fig.~3a or 3b with the red dashed lines). Specifically, we used $0.05\leq\theta\leq0.15$ radians, for all values of $p$, and for both aggregation regimes (BA/MF or DLA/MF). Each $\alpha_{\theta}(p,R)$ follows a power-law behavior as a function of $p$, at fixed $R$, and its initial value, $\alpha_0$, is dictated by the angular scaling of the corresponding BA or DLA model at $p=0$.

\section{Results}

\begin{figure*}[ht!]
\includegraphics[width=\textwidth]{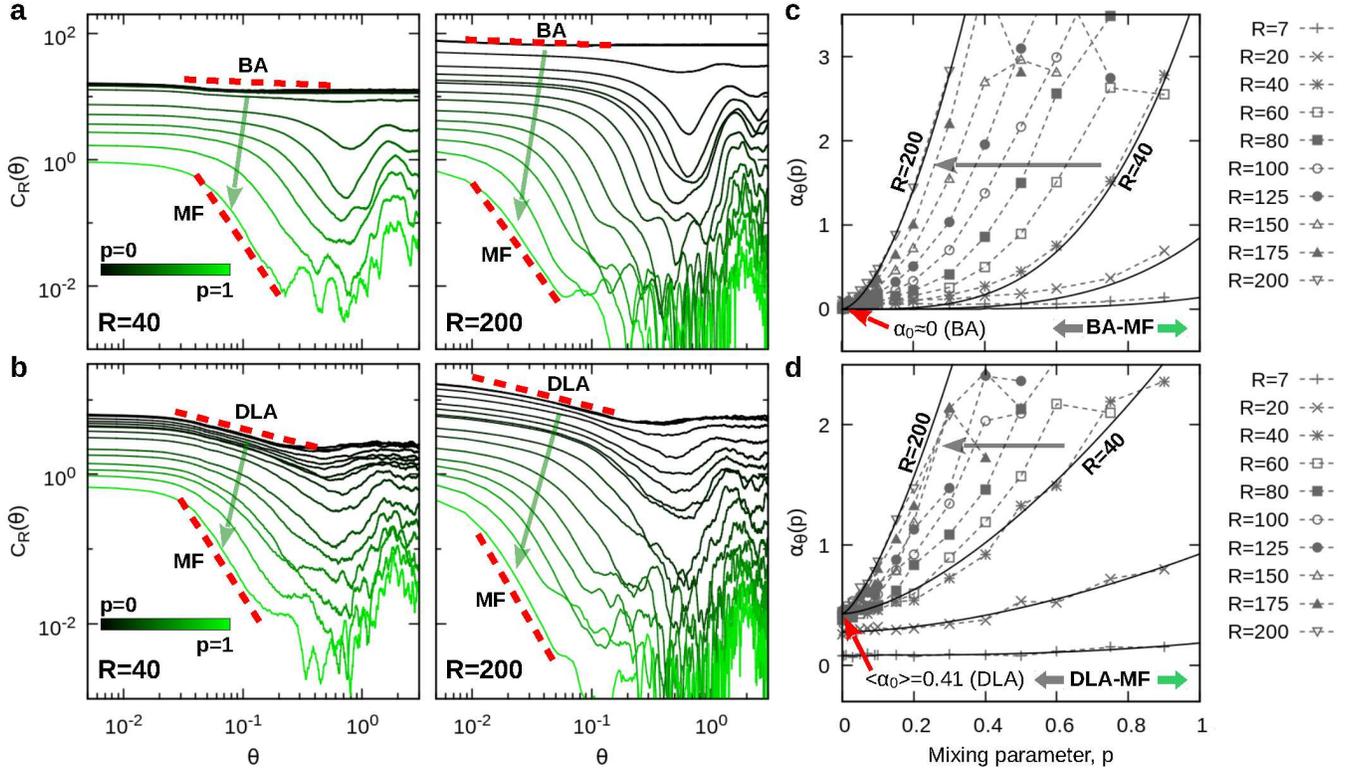}
\caption{\label{fig3}\textbf{Angular correlations and scaling.} (a)-(b) Log-log plots of the angular correlation function, $C_R(\theta)$, of the BA-MF and DLA-MF transitions, respectively, for different control parameters, $p$, and radii, $R$. (c)-(d) Angular scaling exponents, $\alpha_{\theta}(p)$, measured at different radii for the BA-MF and DLA-MF transitions, respectively.}
\end{figure*}

In the following, the angular correlation functions for the BA/DLA-MF transitions are computed for different values of $p$, in order to extract the angular scaling, $\alpha_\theta(p)$. All measurements of $C_R(\theta)$ are performed on an ensemble of 128 clusters, each of which is comprised by $N=1.5\times 10^5$ particles. The maximum radius of analysis is $R=200$. For further details see Methods. For comparison purposes, previously computed fractal dimensions provided by the radial density correlation \cite{nicolas2017}, $D_r(p)$, will be used. In the case of the DBM, all numerical data for $D(\eta)$ is obtained from the literature and listed in detail below.  

\subsection{Angular correlation scaling}

As a starting point, let us consider the general behaviour of $\alpha_\theta$ in the the BA-MF and DLA-MF transitions. For any value of $p$, the angular-density correlation of each cluster obeys, $C_R(\theta)\sim\theta^{-\alpha_\theta}$, for $\theta\ll 1$ (see Figs.~3a and 3b). In these plots, the slope of $C_R(\theta)$ (qualitatively indicated with red dashed lines) not only depends on $p$, but on $R$ as well. Quantitatively, for different fixed values of $R$, the corresponding scaling, $\alpha_{\theta}(p)$, is found to follow the power-law,
\begin{equation}
\label{eq03}
\alpha_{\theta}(p)=\alpha_0+\lambda p^\chi,
\end{equation}
whose steepness increases as $R$ increases (see Figs.~3c and 3d). Here, $R$ should be sufficiently large to avoid finite-size effects. This is better seen by looking at the behaviour of $\alpha_0$ as $p\to 0$ in Figs.~4a and 4b, where the angular scaling for $R\geq 40$ is presented in log-log plots. In these plots, we fitted equation~(\ref{eq03}) to the estimates of $\alpha_{\theta}(p)$ of Figs.~3c and 3d. From these fits, the values of $\lambda$ and $\chi$ were determined. Specific values for this parameters are depicted in the corresponding plots, for measurements performed at $R=40$ (dashed line) and at $R=200$ (solid line). As it can be observed, the description that equation~(\ref{eq03}) gives to $\alpha_\theta$ is valid for any of value of $p$.

In the BA-MF transition (Fig.~4a), $\alpha_0$ converges from $\alpha_0\approx 0.08$, at $R=40$, to $\alpha_0=0$ as $R\gg 1$, in good agreement with the expected theoretical result for a space-filling structure, i.e., $\alpha=0$. In the DLA-MF case (Fig.~4b) we found a consistent $\alpha_0\approx 0.41$, for $p=0$ and $R\geq 40$. On average, for $40\leq R\leq 200$, our measurements rendered $\langle \alpha_0 \rangle=0.41\pm 0.01$, which is in great agreement with previously reported results \cite{meakin1985}, but of course, still different from the expected radial measurement, $\alpha_0=0.29$ \cite{meakin1985,alves2006}.

\begin{figure*}[t!]
\includegraphics[width=\textwidth]{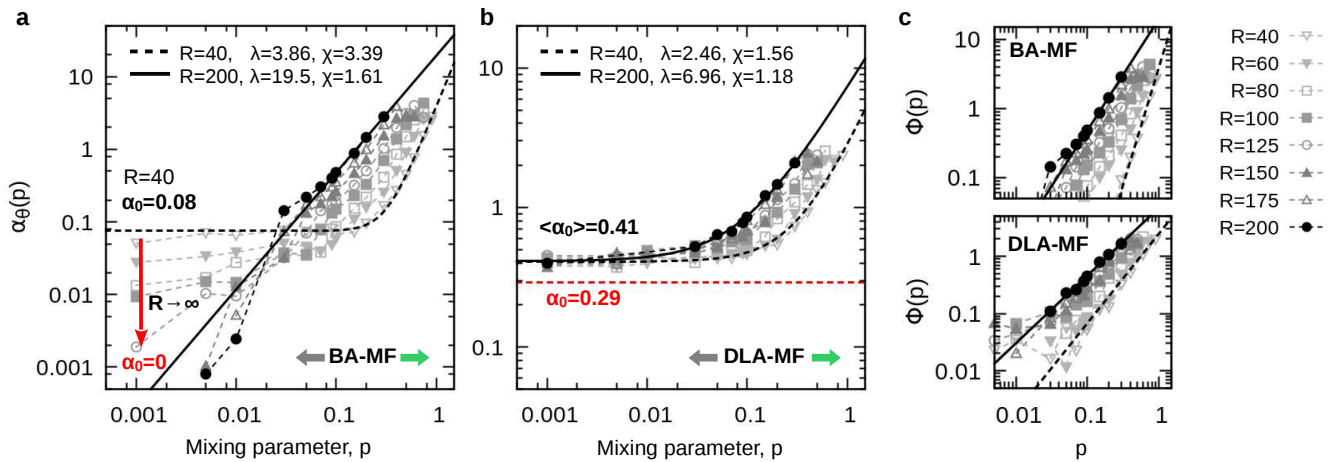}
\caption{\label{fig4}\textbf{Scaling analysis.}  (a)-(b) Angular scaling exponents, $\alpha_{\theta}(p)$, in log-log scale, for $R\geq 40$. (c) Information functions, $\Phi(p)$, for different radii, $R$, of the BA-MF and DLA-MF transitions, as indicated. In plots (a)-(c), black-dashed and black-solid lines are for $R=40$ and $R=200$, respectively, with the parameter values as indicated.}
\end{figure*}

\subsection{Fractality framework}

The power-law behaviour of $\alpha_\theta(p)$, in equation~(\ref{eq03}), can be easily related to the fractal dimension description, $D(D_0,\Phi)$, in equation~(\ref{eq01}). This is done by expanding equation~(\ref{eq01}) into a power-series up to first-order term, $\exp(-\Phi)\approx 1-\Phi$, which yields,
\begin{align}
\label{eq04}
D_\theta(\Phi)=D_0-(D_0-1)\Phi,
\end{align}
with $\Phi=\Phi_0\zeta^\chi$, and $\Phi_0=\Lambda/D_0$. Here, considering that $D_{\theta}=d-\alpha_{\theta}$, equation~(\ref{eq03}) is thus recovered after identifying $\zeta=p$, and $\lambda=(D_0-1)\Phi_0$, while $\chi$ remains as in the original model. In this manner, equation~(\ref{eq04}) allows us to treat the angular scaling as a fractal dimension, $D_\theta(D_0,\Phi)$, defined through the corresponding information function, $\Phi(p)$. This function can be obtained from the previously computed values of $\lambda$ and $\chi$ for each $R$, using equation~(\ref{eq04}), see Fig.~4c.

Under this fractality framework, morphological transitions are described by the radial and angular dimensions of their associated clusters as follows. On the one hand, radial fractal dimensions, $D_r$, described by equation~(\ref{eq01}) \cite{nicolas2017}, will continuously go from $D_0$ at $\zeta=0$ towards $D_r=1$ as $\zeta$ increases. On the other hand, considering that fractal dimensions must satisfy the inequality, $0\leq D\leq d$ \cite{meakin1998book}, we have that the angular fractal dimensions, $D_\theta$, described by equation~(\ref{eq04}), will critically go from $D_0$ at $\zeta=0$, to $D_\theta=0$, at some finite value of $\zeta$. Physically, this point where $D_\theta=0$ corresponds to the scenario in which clusters have fully collapsed to linear structures, $D_r=1$ (i.e., from the lateral or angular direction perspective, these structures are seen as near zero-dimensional or point-like structures). Consequently, there exists a critical or \textsl{cut-off} point, $\zeta_c$, defined as the point where $D_{\theta}=0$, which characterizes the limit where clusters become linear. From equation~(\ref{eq04}), this point is given by,
\begin{equation}
\label{eq05}
\zeta_c=\biggl[\frac{D_0^2}{\Lambda(D_0-1)}\biggr]^{1/\chi}.
\end{equation}
In the radial case, it has been shown \cite{nicolas2017} that the non-linearity of equation~(\ref{eq01}) gives origin to a dynamical \textsl{inflection} point, $\zeta_i$ (the point where the rate of change of $D_r$ becomes a global maximum), given by,
\begin{equation}
\label{eq06}
\zeta_i=\biggl[\frac{D_0(\chi-1)}{\Lambda\chi}\biggr]^{1/\chi}.
\end{equation}
As shown below for the BA/DLA-MF and DBM transitions, these points provide a quantitative measure of the growth-regime changes along the transition. Specifically, $\zeta_i$ characterizes the end of the initial growth-regime (BA or DLA or Eden for the DBM), whereas $\zeta_c$ characterizes the full collapse to linear structures in the final growth-regime (MF or 1D regime in the DBM).

\begin{figure*}[t!]
\includegraphics[width=\textwidth]{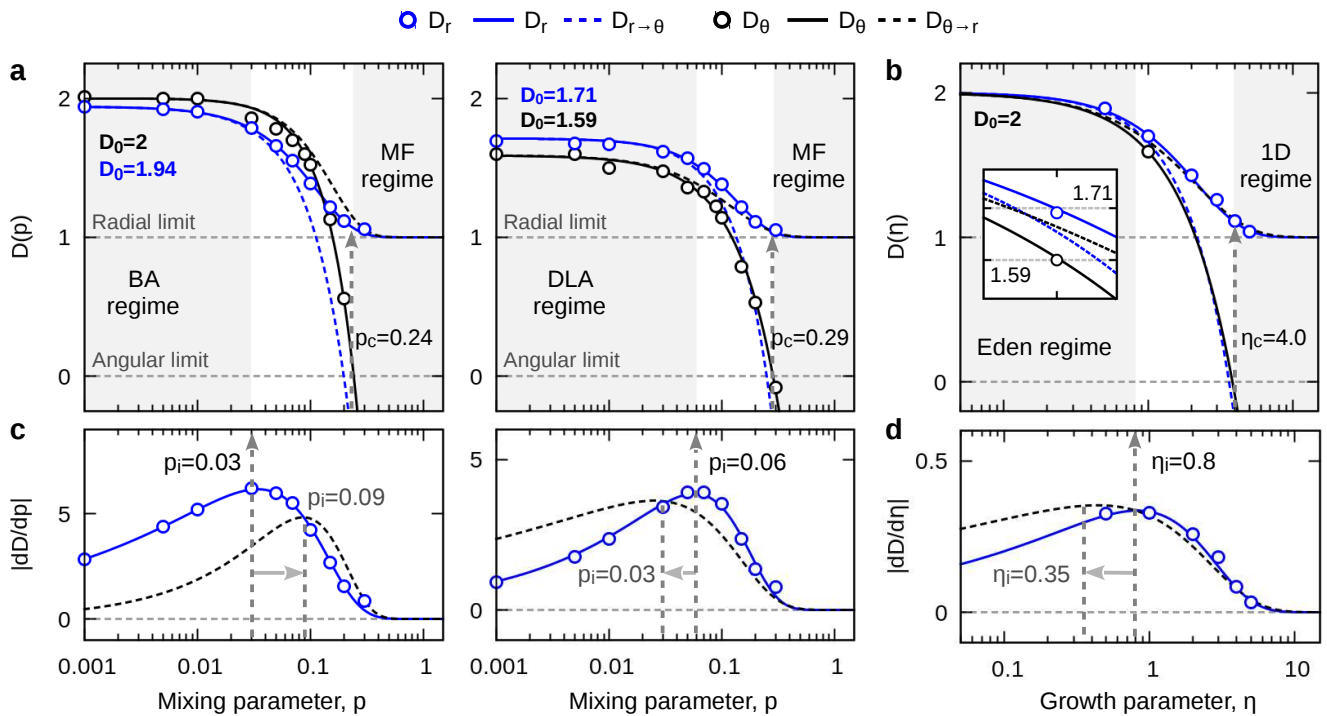}
\caption{\label{fig5}\textbf{Fractality framework.} (a)-(b): Angular and radial fractal dimensions, $D_\theta(\zeta)$ and $D_r(\zeta)$, where $\zeta=\{p,\eta\}$, for the BA/DLA-MF and DBM transitions, respectively. (c)-(d): Derivatives with respect to the control parameter, $|dD/d\zeta|$. In all plots, numerical and analytical results for the angular part are shown in blue, while for the radial part in black. Circles represent the numerical data, solid lines are the corresponding analytical curves, and dashed lines are the analytical curves predicted by the corresponding counterparts (\textit{e.g.}, $D_{r\to\theta}$ is the angular description predicted by the radial part). Dashed arrows are drawn at the inflection and critical points, $\zeta_i$ and $\zeta_c$, respectively. Shaded regions indicate the corresponding growth regimes.
}
\end{figure*}

\subsection{BA-MF and DLA-MF transitions}

In Fig.~5a, the numerical results for the angular and radial scaling of the BA/DLA-MF models are presented under their fractal dimension description. The numerical data for $D_r(p)$ is obtained from previous computations of the scaling of the radial density correlation function \cite{nicolas2017}. The numerical data for $D_\theta(p)$ comes from the scaling $\alpha_\theta(p)$ for $R=200$ (see Fig.~4). The radial fractal dimensions, $D_r(p)$ (shown in black), are perfectly described by the exponential form of equation~(\ref{eq01}), while the angular dimensions, $D_\theta(p)$ (shown in blue), are described by the power law form of equation~(\ref{eq04}). For this analytical description, the corresponding information functions were numerically obtained from equation~(\ref{eq04}), by plotting $\Phi(p)=-[D_\theta-D_0]/[D_0-1]$, followed by the use of the relation $\Phi(p)=\Phi_0p^\chi$, as a fitting function (see Fig.~4c). The specific numerical values of $\Lambda$ and $\chi$ are presented in Table I. According to the fractality framework, from $D_\theta(p)$ one obtains $p_c$, this is, the point of full collapse to linear structures at the MF regime, while $p_i$ is obtained from the inflection condition of $D_r(p)$, that is, the point where the initial growth-regime (BA or DLA) ends and the rate of change of $D_r$ becomes a maximum (see Fig.~5c). The numerical values of $p_i$ and $p_c$ are presented in Table I. 

\subsection{DBM transition}

In Fig.~5b, the numerical data for the radial and angular fractal dimensions of the DBM are presented. Here, the radial fractal dimensions $D_r(\eta)$ are obtained as the average of the reliable numerical results reported in the literature \cite{pietronero1988,somfai2004,tolman1989b,amitrano1989,hastings2001} (see Table II). For the angular part, based on the previous results for the BA/DLA-MF transitions, equations~(\ref{eq04}) and (\ref{eq05}) are used to retrieve the complete angular information. First, we used $D_\theta=1.59$, at $\eta=1$ (the DLA point) with equation~(\ref{eq04}) to solve for $\Lambda$, leading to $\Lambda=0.82$. Secondly, we used $D_\theta=0$ at $\eta_c=4$ (which corresponds to $D_r=1$ for the radial part) with equation~(\ref{eq05}), to solve for $\chi$, leading to $\chi=1.14$. From equation~(\ref{eq06}), we computed $\eta_i=0.8$, which is the point of maximum rate of change in $D_r(\eta)$ and the end of the Eden regime (see Fig.~5d). These numerical results are shown in Table I.

\section{Discussion}

\begin{figure*}[t!]
\includegraphics[width=\textwidth]{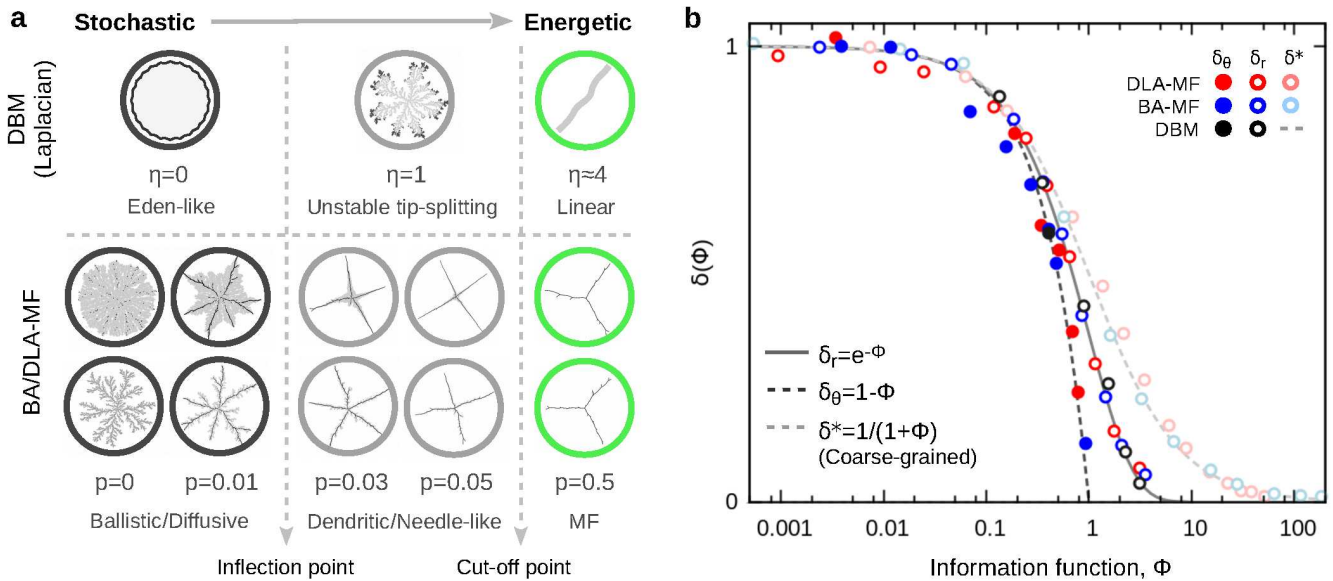}
\caption{\label{fig6}\textbf{Unified fractality framework.} (a) Dynamical and morphological growth regimes under the $D(D_0\Phi)$ framework. (b) Universal description, $\delta(\Phi)$, in terms of the information function ($\Phi$).}
\end{figure*}

The following discussion is centred on four points: (I) the validity limits of the radial/angular equivalence ($D_r=D_\theta$); (II) the levels of predictability, this is, the extent to which one can infer the radial/angular descriptions from the previous knowledge of the other one; (III) the role of both fractal dimensions as non-thermal order parameters; (IV) the unified fractality framework that the general dimensionality function $D(D_0,\Phi)$ provides to the angular- and radial-density correlation scaling results. This includes previous radius of gyration and mean-field results.

\textbf{I. Validity of the radial/angular equivalence.} The exponential form in equation~(\ref{eq01}) and power-law in equation~(\ref{eq04}) give an excellent description to the numerical results for the radial and angular dimensions, $D_r$ and $D_\theta$, respectively, as function of the corresponding control parameter (see Fig.~5a and 5b). From these results, one can observe that the radial and angular descriptions are equivalent only under certain conditions. 

In the BA-MF case (Fig.~5a), the angular and radial measurements are in good agreement as $p\to 0$. On the radial part, $D_0\approx 1.94$ is close to the usual measurements found for clusters of this size ($N\sim 10^5$) \cite{alves2006}, nevertheless, without further controversy, it is expected to converge to $D_0=2$ as $N\to\infty$ \cite{meakin1998book}. On the other hand, the angular measurement is consistent with the expected theoretical value ($D_0=2$). Hence, the radial/angular equivalence holds. However, this is no longer true for the DLA-MF case, in which $D_0=1.71$ is the dimension for the radial part but $D_0=1.59$ is the consistent result found for the angular part \cite{meakin1985,alves2006}. This discrepancy cannot be accounted to finite-size effects since the value for the angular $\alpha_0$ remains constant for different measurements with $R\geq 40$ (and up to $R=200$ in this analysis) (see Fig.~4b). Rather, this issue points towards two different scaling mechanisms of the angular and radial parts in DLA that arise due to the effects of the inherent anisotropies developed during DLA growth \cite{meakin1985,alves2006}. 

Indeed, these observations come together and are better understood by looking at the scaling dynamics of the DBM (see Fig.~5b). In this case, the radial and angular measurements have an exact match of $D_0=2$ at $\eta=0$ (similar to $p=0$ in the BA-MF transition), but they eventually diverge for $\eta>0$, as growth anisotropies start to develop. In particular, at $\eta=1$ (equal to the DLA-MF transition at $p=0$), the difference of the radial and angular measurements is then expected due to the anisotropies presented in DLA fractal (not finite-size effects in the measurements). Additionally, this difference in the angular and radial measurements would also predicted by equation~(\ref{eq01}), since by setting $D_0=2$, this equation indicates that the breaking of the radial/angular equivalence for $\eta>0$ arises from the approximated nature of $D_\theta$ in equation~(\ref{eq04}), which was obtained from the first-order approximation of the general exponential form of $D_r$ in equation~(\ref{eq01}). 

Therefore, as supported by the numerical and analytical results, we can conclude that the often used radial/angular equivalence, $D_r=D_\theta$, is only valid for homogeneous compact clusters (such as BA or Eden in the DBM). Asides from this, unavoidable growth anisotropies (as subtle as those presented in DLA) render this equality invalid. Under these circumstances, the interpretation of $D_\theta$ as a formal fractal dimension \cite{sander2011} must treated with caution (e.g. $D_\theta=0$ for a linear structure which trivially has a dimension $D=1$). At best, near the homogeneous compact growth regime (see Fig.~5), the angular fractal dimension is a good approximation to the real dimension. 

\begin{table*}[ht!]
\caption{\textbf{Numerical results.} Elements marked with a star are the values predicted by the corresponding counterparts.}
\begin{ruledtabular} 
\begin{tabular}{llllllllll}
Transition & $\zeta$ & $D$  & $D_0$ & $\Lambda$ & $\chi$ & $\zeta_i$ & $D_i$ &  $\zeta_c$ & $D_c$ \\
\hline
BA-MF 	&$p$ & $D_r$	& $1.94\pm 0.01$ 	& $31.8\pm 2.6$	& $1.28\pm 0.03$	& 0.034	 & 1.76   & 0.20* & 1.12* \\
& & $D_{\theta}$	& $2$	& $39.1\pm 1.0$	& $1.61\pm 0.04$	& 0.086* & 1.68* &	 0.24  &  1.06 \\
DLA-MF 	& $p$&$D_r$	& $1.71$ 			& $29.3\pm 4.9$	& $1.42\pm 0.08$	& 0.057	 & 1.53   & 0.25*	& 1.06* \\
& & $D_{\theta}$	& $1.59\pm 0.01$ & $18.8\pm 0.5$	& $1.18\pm 0.06$	& 0.025* & 1.51* & 0.29   &  1.04 \\
DBM 	&$\eta$ & $D_r$	& $2$ 	& $0.69$	& $1.36\pm 0.02$	& 0.82	 & 1.77   & 3.6* & 1.14* \\
& & $D_{\theta}$	& $2$	& $0.82$	& $1.14$			& 0.35*  & 1.88* & 4.0  &  1.10 \\
\end{tabular}
\end{ruledtabular} 
\end{table*}

\begin{table*}[ht!]
\caption{\textbf{Compilation of DBM results.} The average fractal dimension $\langle D(\eta)\rangle$, obtained from the average of reliable numerical results reported in the literature. Errors in measurements are shown when available.}
\begin{ruledtabular} 
\begin{tabular}{llllllll}
Data & $\eta=0.5$ & $\eta=1$ & $\eta=2$ & $\eta=3$ & $\eta=4$ & $\eta=5$ \\
\hline
Amitrano \cite{amitrano1989} & $1.86$ & $1.69$ & $1.43$ & $1.26$ & $1.16$ & $1.07$ \\
Pietronero, \textit{et al.} \cite{pietronero1988} & $1.92$ & $1.70$ & $1.43$ & & & \\
Somfai, \textit{et al.} \cite{somfai2004} & & $1.71$ & $1.42$ & $1.23$ & & \\
Tolman \& Meakin \cite{tolman1989b} & & & $1.408\pm 0.006$ & $1.292\pm 0.003$ & & \\
Hastings \cite{hastings2001} & & & $1.44\pm 0.01$ &	$1.26\pm 0.01$ &	$1.09\pm 0.03$ & $1.03\pm 0.02$ \\
$\langle D(\eta)\rangle$	&$1.89\pm 0.05$	&$1.70\pm 0.01$	&$1.43\pm 0.02$	&$1.26\pm 0.02$	&$1.11\pm 0.04$	&$1.04\pm 0.03$\\
$D_r(\eta)$					&$1.87$		&$1.71$		&$1.41$		&$1.22$		&$1.10$		&$1.05$ \\
$D_{\theta\to r}(\eta)$		&$1.83$		&$1.66$		&$1.41$		&$1.24$		&$1.14$		&$1.08$ \\
$D_{MF}(\eta)$ \cite{muthukumar1983,tokuyama1984,matsushita1986}		&$1.80$		&$1.66$		&$1.50$	&$1.40$	&$1.33$	&$1.29$ \\
\end{tabular}
\end{ruledtabular}
\end{table*}

\textbf{II. Levels of predictability.} Since $D_\theta$ is given by equation~(\ref{eq04}), which was derived as a first-order approximation of the more general form for $D_r$ in equation~(\ref{eq01}), let us now explore the extent to which the radial/angular descriptions can be inferred from the previous knowledge of the other one. 

To this end, let us denote as $D_{r\to\theta}$ to the predicted values of $D_\theta$ given the knowledge of the radial counterpart $D_r$. This is, given that all numerical values for $D_0$, $\Lambda$, and $\chi$ are known for the radial dimensions $D_r$ in equation~(\ref{eq01}), let us use the same numerical values for the angular dimensions $D_{r\to\theta}$ in equation~(\ref{eq04}). Similarly, $D_{\theta\to r}$ denotes the predicted values of $D_r$ given the knowledge of the angular counterpart $D_\theta$. This is, given that all numerical values for $D_0$, $\Lambda$, and $\chi$ are known for the angular dimensions $D_\theta$ in equation~(\ref{eq04}), let us use the same numerical values for the radial dimensions $D_{\theta\to r}$ in equation~(\ref{eq01}).

Considering this, we found that there are no good levels of prediction or inference for any of the BA/DLA-MF cases (see Fig.~5a). This is, the inferred $D_{r\to\theta}$ is not equal to the measured $D_\theta$, nor $D_{\theta\to r}$ is equal to the measured $D_r$. However, certain level of prediction seems possible for the DBM (see Fig.~5b). For example, if one attempts to recover the radial scaling from the angular part, then, $D_r$ in equation~(\ref{eq01}) must be used together with the parameters of the angular part, $\Lambda=0.82$ and $\chi=1.14$ (that in equation~(\ref{eq04}) give $D_\theta=1.59$ at $\eta=1$). This gives as a result $D_{\theta\to r}=1.66$, which is a very common and quite close result to the accepted $D_r=1.71$; \textit{vice~versa}, recovering the angular dimensions from the radial part using the radial scaling parameters $\Lambda=0.69$ and $\chi=1.36$ (that in equation~(\ref{eq01}) give $D_r=1.71$ at $\eta=1$) yields $D_{r\to \theta}=1.66$, which again, is a commonly reported result, although not as close as to the common $D_\theta=1.59$ (see Table II). 

One possible explanation as to the impossibility of prediction in the BA/DLA-MF cases can be attributed to the growth characteristics of their corresponding clusters. Specifically, these clusters exhibit multi-scaling features, i.e., scaling that depends on the scale of measurement (self-similarity breaking) \cite{nicolas2017}. In the DBM case, clusters are self-similar and notably, differences between inferred and real measurements are minor. Therein, from these results we can conclude that certain levels of radial/angular prediction or inference, from the results of one method to the other, are possible just in the case of morphological transitions characterized by clusters with well-defined scaling (such as in the DBM). 

\textbf{III. Order parameters.} In terms of their morphological features, these transitions go from dense or dendritic branching (disordered states), to linear clusters (ordered state). In terms of their corresponding dynamical growth regimes, they go from the stochastic regime, with uniform (BA, Eden) or unstable tip-splitting (DLA) growth, towards the energetic regime, with highly anisotropic growth (MF or $\eta\geq 4$ in the DBM) (see Fig.~6a). 

In general, the (radial) fractal dimension has been used as a non-thermal order parameter for these systems \cite{hastings2001}, however, the characterization that it provides to the dynamical growth regimes is limited. Under the $D(D_0,\Phi)$ model used in this analysis, a complete set of morphological measures is provided by $D_r$ and $D_\theta$ in equations~(\ref{eq01}) and (\ref{eq04}), respectively, along with the inflection and cut-off points, $\zeta_i$ and $\zeta_c$ in equations~(\ref{eq05}) and (\ref{eq06}), respectively. As seen in Fig.~5c and 5d, $\zeta_i$ defines the end of the stochastic regime (BA, DLA or Eden) at the point of maximum rate of change of $D_r$, whereas $\zeta_c$ clearly defines the highly energetic regime (MF or $\eta\geq 4$ in the DBM) at the point of full collapse to the linear regime, in the $D_\theta=0$ limit (see Table I). 

Notably, these particular ways to define the order/disorder growth regimes have a strong resemblance (although they are not formally equal) to ordinary second- and first-order phase transitions. Hence, in this morphological transitions where $D(\zeta)$ might be considered a non-thermal order parameter with $\zeta$ an ``effective temperature'', we have that $D_r$ behaves like a second-order transitions (with transition point at $\zeta_i$) while $D_\theta$ like a first-order one (with transition point at $\zeta_c$).  

\textbf{IV. Unified fractality framework.} The $D(D_0,\Phi)$ model has been successfully applied to the analytical description of other quantification methods \cite{nicolas2017} and quite recently, to the DBM in any Euclidean dimension \cite{nicolas2019}. For example, it has been applied to coarse-grained descriptions, as that given by the radius of gyration, $R_g(N)\sim N^\beta$, where $N$ is the number of particles in a cluster. Here, the fractal dimensions are obtained from $D=1/\beta$, and described by a first-order approximation of equation~(\ref{eq01}), in the form given by $\exp(-\Phi)\approx 1/(1+\Phi)$, which leads to $D^*(D_0,\Phi)=(D_0+\Phi)/(1+\Phi)$. This model has also been applied to the well-known DBM mean-field equation \cite{muthukumar1983,tokuyama1984,matsushita1986}, $D_{MF}(\eta)=(d^2+\eta)/(d+\eta)$, that belongs to the same approximation as the radius of gyration, with $D_0=d$ and $\Lambda=\chi=1$. Thus, considering that the fractal dimensions obtained from the scaling of the radial and angular density correlations can also be treated under the $D(D_0,\Phi)$ description, the possibility of a unified analytical framework for fractal/non-fractal morphological transitions is then evident. 

This framework is neatly summarized under the transformation given by $\delta=(D-1)/(D_0-1)$, that when applied to $D_r$ in equation~(\ref{eq01}) (radial-density correlation), $D_\theta$ in equation~(\ref{eq04}) (angular-density correlation), and $D^*$ (coarse-grained descriptions such as radius of gyration and mean-field theories), we respectively have,
\begin{align}
\label{eq07}
\delta_r(\Phi)&=e^{-\Phi}, \nonumber \\
\delta_\theta(\Phi)&=1-\Phi, \\
\delta^*(\Phi)&=1/(1+\Phi). \nonumber
\end{align}
In Fig.~6, the $\delta$-transformation is applied to the data used in Fig.~5 (as well as to previous results for the scaling of the radius of gyration of the BA/DLA-MF transitions \cite{nicolas2017} and the mean-field equation of the DBM) and plotted as function of their associated information-functions. This transformation causes all the data to collapse into single curves according to their corresponding description (radial, angular, or coarse-grained). In this way, equations~(\ref{eq07}) show that the mathematical description of these systems not only is quite simple, but more general than previously found \cite{nicolas2017, nicolas2019}. In addition, they demonstrate the existence of a universal description that, in particular, is independent of the fractal dimension of the initial configuration, the geometrical symmetry-breaking process that drives the transition, and the Euclidean dimension of the embedding space.

\section{Conclusions}

From the previous discussion, the main conclusions per point are: 

I. The often used radial/angular equivalence, $D_r=D_\theta$, is only valid for homogeneous compact clusters (such as BA or Eden in the DBM). Asides from this, unavoidable growth anisotropies (as subtle as those presented in DLA) render this equality invalid. Under these circumstances, the interpretation of $D_\theta$ as a formal fractal dimension \cite{sander2011} must be treated with caution (e.g. $D_\theta=0$ for a linear structure which trivially has a dimension $D=1$). At its best, near the homogeneous compact growth regime (see Fig.~5), the angular fractal dimension is a good approximation to the real dimension.

II. Certain levels of radial/angular prediction or inference, from the results of one method to the other, are possible just in the case of morphological transitions characterized by clusters with well-defined scaling (such as in the DBM). 

III. In this morphological transitions where $D(\zeta)$ might be considered a non-thermal order parameter with $\zeta$ an ``effective temperature'', $D_r$ behaves like a second-order transitions (with transition point at $\zeta_i$) while $D_\theta$ like a first-order one (with transition point at $\zeta_c$). 

IV. Under the framework provided by $D(D_0,\Phi)$ in equation~(\ref{eq01}), the numerical and analytical results from different methodologies come together under the transformation given by $\delta=(D-1)/(D_0-1)$, that when applied to $D_r$ in equation~(\ref{eq01}) (radial-density correlation), $D_\theta$ in equation~(\ref{eq04}) (angular-density correlation), and $D^*$ (coarse-grained descriptions such as radius of gyration and mean-field theories), become equations~(\ref{eq07}), respectively. These equations demonstrate the existence of a universal description, this is, independent of the fractal dimension of the initial configuration, the geometrical symmetry-breaking process that drives the transition, and the Euclidean dimension of the embedding space (see Fig.~6). 

As final remarks, we must point out that fractal/non-fractal morphological transitions are found in all sorts of natural and artificial systems, and they are not an exclusive feature of particle aggregation models. Nonetheless, these simple models are a powerful asset that allow for the systematic study of diverse structural and structural-dependent processes (over a wide range of spatial scales and organic-like morphologies), which might be more complicated under other approaches. In this work, we showed that, despite the morphological complexity of these systems and the different quantification methodologies, the construction of a unified scaling analysis framework is possible. We are positive that the models and results presented here will find creative applications in current scientific and technological areas dealing with morphodynamics in complex systems research.

\section*{Acknowledgments} 

Authors acknowledge partial financial support by CONACyT through the Soft Matter Network and VIEP-BUAP, grants CAREJ-EXC17-G and SOAJ-NAT17-I.

\end{document}